\documentclass[fleqn,twoside]{article}
\usepackage{espcrc2}
\usepackage{graphics}

\title{
Search for Lepton Flavor-Violating $\tau \to \mu \gamma$ Decay
}

\author{
K.Inami
\address[Nagoya]{Department of Physics, Nagoya University, \\
Furo-cho, Chikusa-ku, Nagoya, 464-8602, Japan},
T.Hokuue\addressmark[Nagoya],
T.Ohshima\addressmark[Nagoya]
for the Belle collaboration
}

\begin{document}

\begin{abstract}
We search for the lepton flavor-violating $\tau \to \mu \gamma$ decay
using 29.7 million $\tau$ pairs accumulated by the Belle experiment.
The main background sources are found to be
$\tau$ pairs with $\tau \to \mu \nu \nu$ decay
and radiative $\mu\mu$ events.
One event is observed in the signal region, while $2.5\pm0.6$ background events 
are expected. A preliminary upper limit
$Br(\tau \to \mu \gamma) < 6 \times 10^{-7}$
at the 90\% confidence limit is obtained.
\end{abstract}

\maketitle

\section{Introduction}

Charged lepton flavor-violating decays, such as $\tau \to \mu \gamma$,
$e \gamma$, $\mu\mu\mu$ and $\mu \to e \gamma$,
are forbidden in the Standard Model (or highly suppressed
even if we consider neutrino mixing).
However, new physics beyond the Standard Model
allows lepton flavor violation (LFV) decays.
In supersymmetric models, left-right symmetric models and others~\cite{SUSY},
the branching ratio of the $\tau \to \mu \gamma$ decay,
Br($\tau \to \mu\gamma$), 
is predicted to be $10^{5\mbox{-}6}$ times 
higher than Br($\mu \to e \gamma$) because of the large $\tau$ mass.
Also, some SUSY models predict Br $\sim$ O($10^{-7}$),
which can be reachable by B-factory experiments of high luminosity.

The current upper limit of Br($\tau \to \mu \gamma$)
is $1.1 \times 10^{-6}$ (90\%CL), measured by the CLEO experiment~\cite{CLEO}. 
Our first analysis with 10 fb$^{-1}$ of data resulted in
a slightly better upper limit of $1 \times 10^{-6}$ (90\%CL)~\cite{EPS01}.
Also, the BaBar group recently reported an upper limit of
$2.0 \times 10^{-6}$ (90\%CL) using 56 million $\tau$ pairs 
at the ICHEP2002 conference~\cite{ICHEP02}.

Data is accumulated with the Belle detector~\cite{Belle} 
at the KEKB accelerator~\cite{KEKB}.
KEKB is an asymmetric $e^+ e^-$ collider with a center-of-mass energy
of 10.58 GeV. Its current peak luminosity is
$7.4 \times 10^{33} {\rm cm}^{-2}{\rm s}^{-1}$ and
the total integrated luminosity amounts to 89.6 fb$^{-1}$ 
as of summer, 2002.
Here we present an analysis using 32.6 fb$^{-1}$ of data, corresponding to 
29.7 million $\tau$ pairs.

\section{Event selection}

\begin{figure*}[t]
\centerline{\resizebox{5cm}{5cm}{\includegraphics{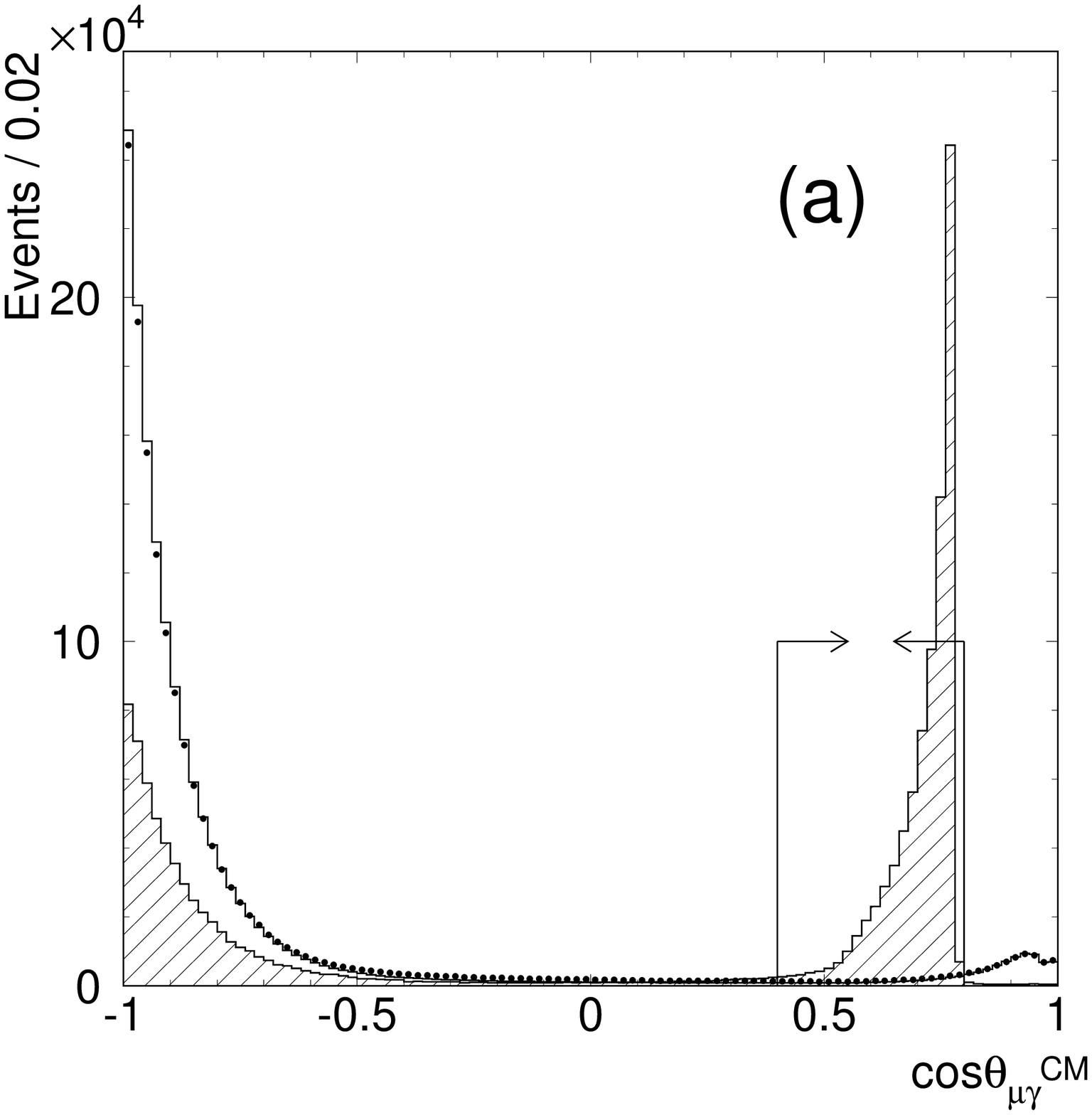}}~~~
            \resizebox{5cm}{5cm}{\includegraphics{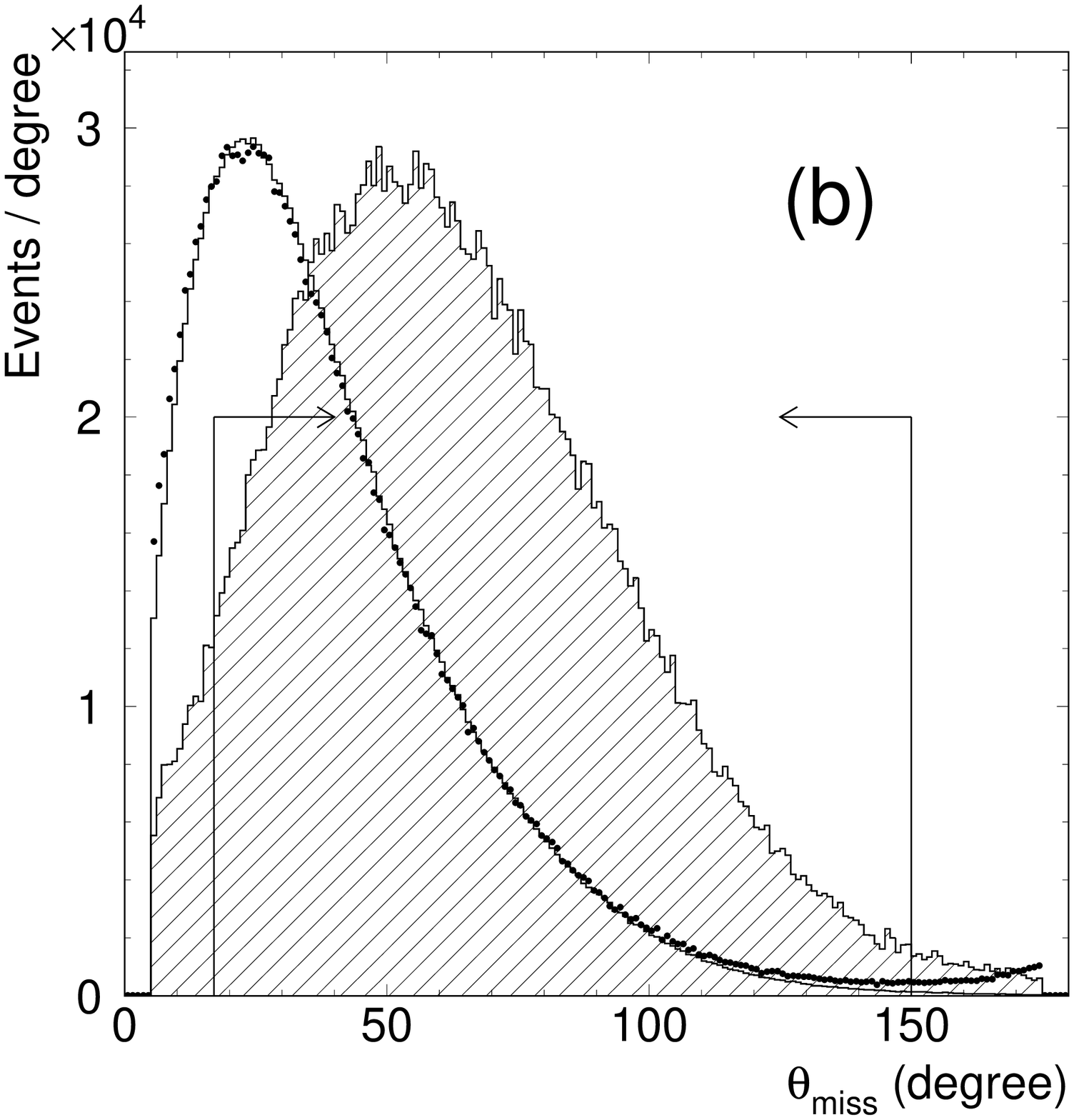}}~~~
            \resizebox{5cm}{5cm}{\includegraphics{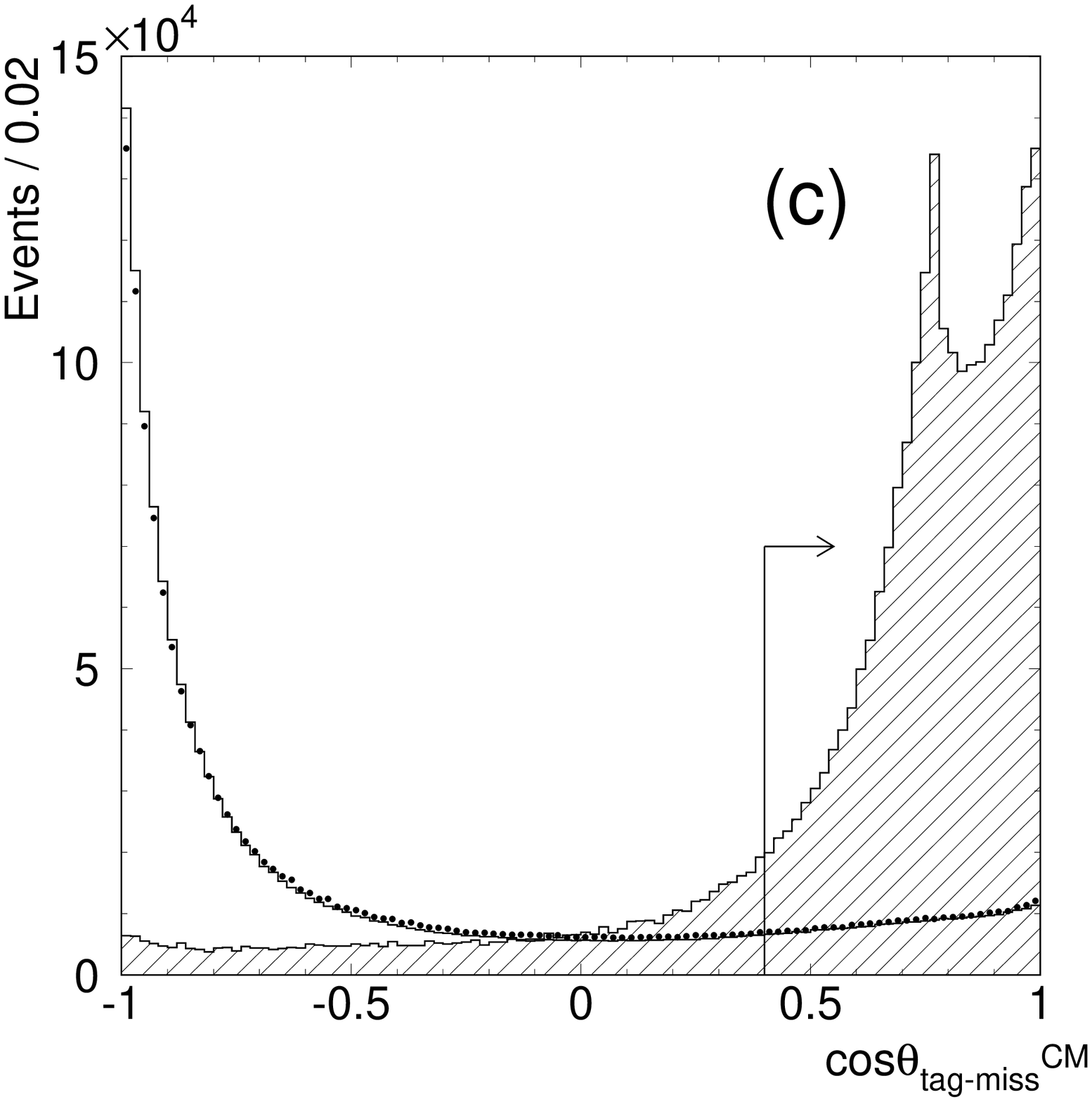}} }
\caption{Comparison of some kinematical distributions among data (dots),
$\tau$ pair background MC (open histogram) and signal MC (hatched histogram).
The arrows indicate the selection criteria.
}
\label{cuts}
\end{figure*}
\begin{figure*}
\centerline{\resizebox{5cm}{5cm}{\includegraphics{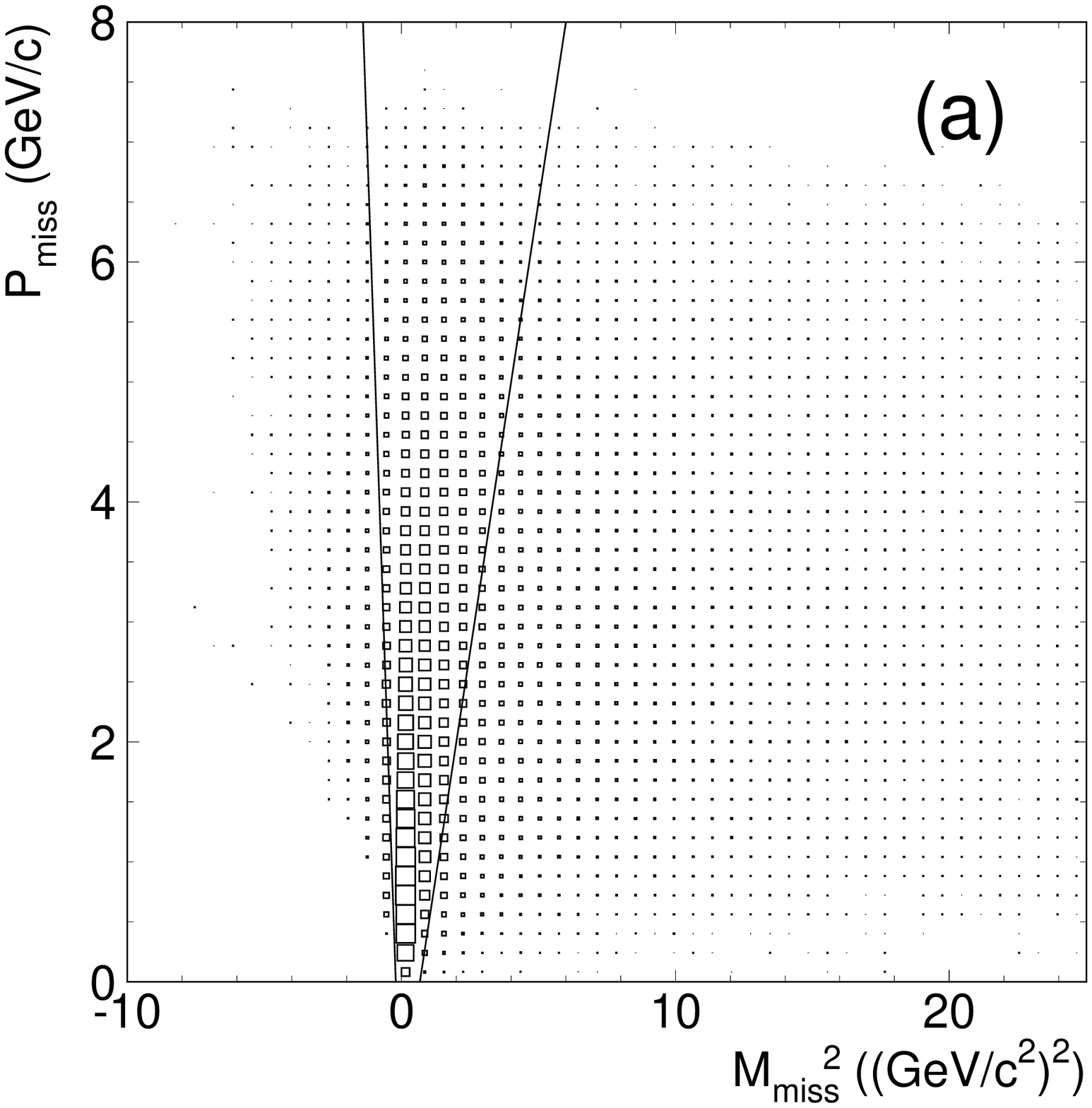}}~~~~~~~~
            \resizebox{5cm}{5cm}{\includegraphics{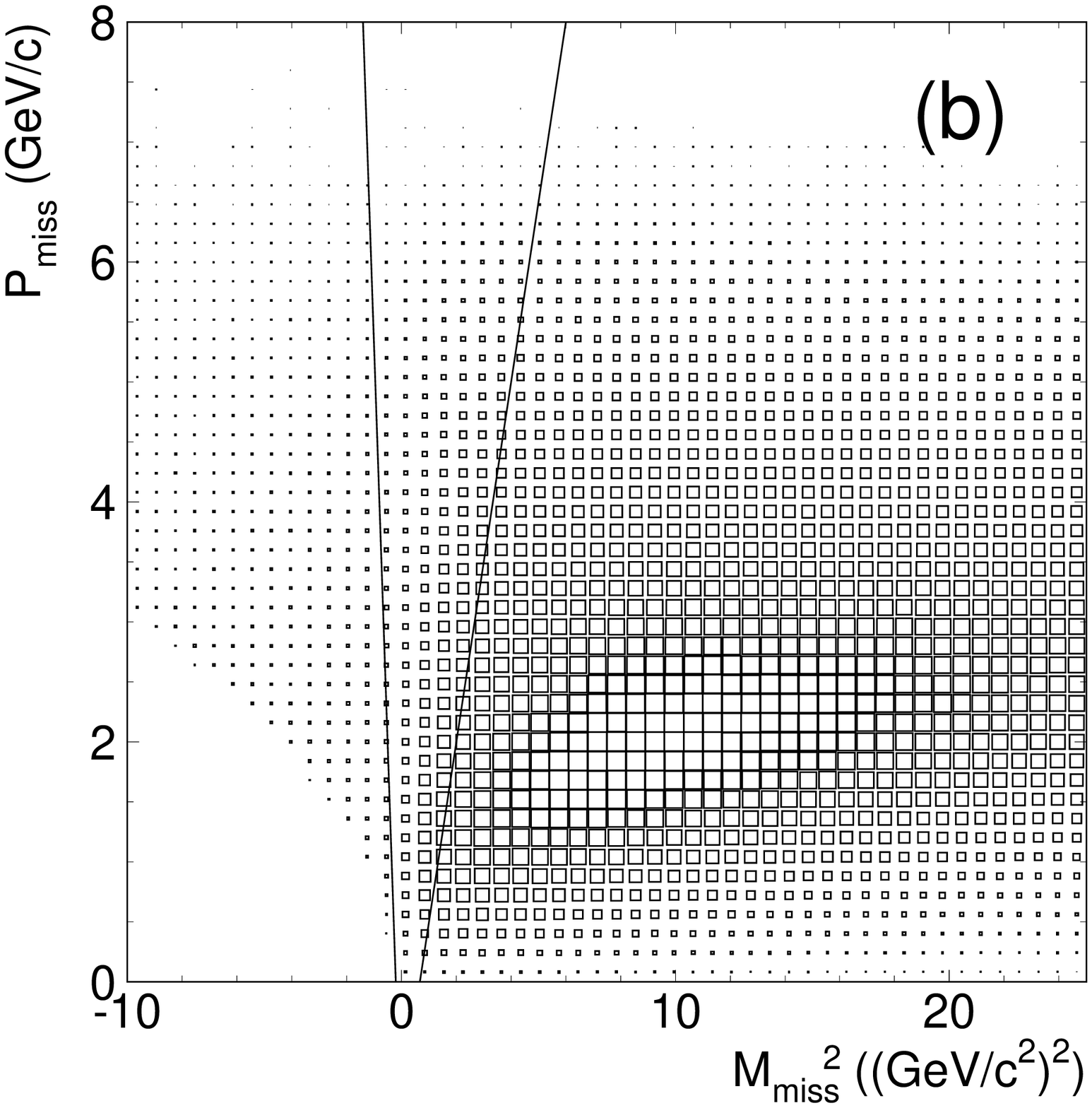}}}
\caption{Correlations between $p_{\rm miss}$ and $m^2_{\rm miss}$ 
for (a) the signal MC and (b) the $\tau$-pair background MC.
The lines show the selection boundaries applied in this analysis.
}
\label{cut4}
\end{figure*}

In order to determine the selection criteria, 
a Monte Carlo (MC) simulation is performed using the KORALB generator 
\cite{KORALB} for $\tau$-pair samples
and a QQ generator \cite{QQ} for other physics processes, as noted below.
GEANT \cite{GEANT} is used for the full detector simulation.
We prepare two kinds of $\tau$ pair samples: signal and 
$\tau$-pair background, and several kinds of background samples: 
${\rm B^0\overline{B}^0}$, ${\rm B^+B^-}$, 
continuum, Bhabha, $\mu\mu$ and two-photon processes of ${\rm eeee}$, 
${\rm ee\mu\mu}$, ${\rm eeu\bar{u}/d\bar{d}}$, ${\rm ees\bar{s}}$ 
and ${\rm eec\bar{c}}$. 
In the signal MC samples, one $\tau$ decays to $\mu\gamma$ and 
the other $\tau$ decays generically.
The angular distribution of two-body $\tau \to \mu \gamma$ decay
is assumed to be uniform in the $\tau$'s rest frame.

A candidate event is required to be \\
$~~~~(\mu\gamma)+[{\rm (a~ charged~ particle~ but~ not~ \mu)}$ \\
$~~~~~~~~~~~~~~~~~~~~~~~~~~~~~~~~~+{\rm (\geq0\gamma)+missing}],$ \\
which we hereafter denote as ``$\mu^{\rm not}\mu$'' events. 
We therefore select events with ``2 charged-tracks and $\geq$1 $\gamma$''
with the expected accompanying missing momentum.
Below, kinematical variables are defined in the laboratory frame,
while those defined in the center-of-mass frame of the system are indicated
with an index ``CM''.
Two oppositely charged particles are required to have a transverse momentum
of $p_t > 0.1$ GeV/$c$ and a momentum of $p^{\rm CM} < 4.5$ GeV/$c$ 
to reject Bhabha and $\mu\mu$ background events.
For the signal, a muon track is required to satisfy
$-0.819<\cos \theta_\mu<0.906$ and $p_\mu > 1.0$ GeV/$c$, and
to have a $\mu$-probability of more than 0.90
from the K$_{\rm L}$/muon detector (KLM).
The signal photon is selected by requiring
$-0.866<\cos \theta_\gamma<0.956$ and $E_\gamma > 0.5$ GeV.
The charged particle on the tag side is required to not be a muon
within $-0.819< \cos \theta_{\rm tag}<0.906$.
In order to remove the background from radiative $\mu\mu$, two-photon
and other non-$\tau$ processes, we apply the following criteria.
The opening angle between the two charged particles be greater than $90^\circ$,
and for the missing particle, $p_{\rm miss} > 0.4$ GeV/$c$ and 
$17^\circ < \theta_{\rm miss} < 150^\circ$.

To remove the $\tau$-pair background,
we demand the opening angle between the $\mu$ and $\gamma$ of 
the signal candidate 
to be $0.4< \cos \theta^{\rm CM}_{\mu \gamma} <0.8$
and the opening angle between 
the charged particle in tag side and the missing momentum 
to be $\cos \theta^{\rm CM}_{\rm tag-miss}>0.4$.
The distributions for these variables are shown in Fig.~\ref{cuts}.
A further selection is applied for
the missing momentum ($p_{\rm miss}$)
and the missing mass squared ($m^2_{\rm miss}$) as
$p_{\rm miss} > -5 m^2_{\rm miss}-1$ and 
$p_{\rm miss} > 1.5 m^2_{\rm miss}-1$,
as shown in Fig.~\ref{cut4}.
With these selections, the signal efficiency is evaluated to be 12.8\% by MC and
the remaining rate of background is $5.7 \times 10^{-6}$ for 
$\tau$-pairs and $2 \times 10^{-6}$ for the radiative $\mu\mu$.

\begin{figure}
\centerline{\resizebox{7cm}{7cm}{\includegraphics{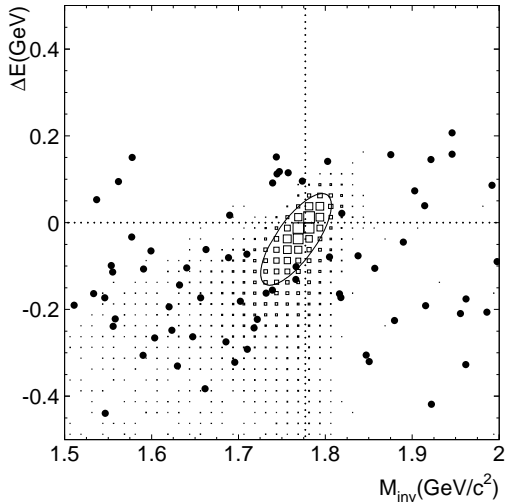}}}
\caption{$M_{\rm inv}$-vs-$\Delta E$ plot for events which passed 
all selections.
The data events are indicated by dots and the MC events by boxes.
The ellipse shows the signal region.}
\label{2d_data}
\end{figure}

The signal region on the $M_{\rm inv}$-$\Delta E$ plane
is determined using the signal MC while assuming a uniform
background distribution around the signal peak, where $M_{\rm inv}$ is the 
invariant mass of the $\mu$-$\gamma$ system and $\Delta E$ is the energy
difference between the $\mu$-$\gamma$ system and the beam energy 
in the CM frame.
Although we used a box shape to define the signal region in previous analysis,
we now employ an elliptic shape in order to enhance the sensitivity,
because the signal distribution has a tail 
due to the initial-state radiation and photon energy leakage
from the CsI photon detector.
Fig.~\ref{2d_data} shows an ellipse with the highest signal-to-background
ratio for a total detection efficiency of 9.0\%.

With all selections, 69 events remain
within the regions called ``Area'',
defined to be $1.5<M_{\rm inv}<2.0$ GeV/$c^2$ and 
$-0.5<\Delta E<0.5$ GeV.
One event is observed in the signal region, as plotted 
in Fig.~\ref{2d_data}.

\section{Background estimate}

It is not straightforward to use a sideband to estimate
the background in the signal region, since it largely depends on the 
sideband allocation for small samples, especially 
in the case of a structural distribution.

\begin{figure*}[t]
\centerline{\resizebox{5cm}{5cm}{\includegraphics{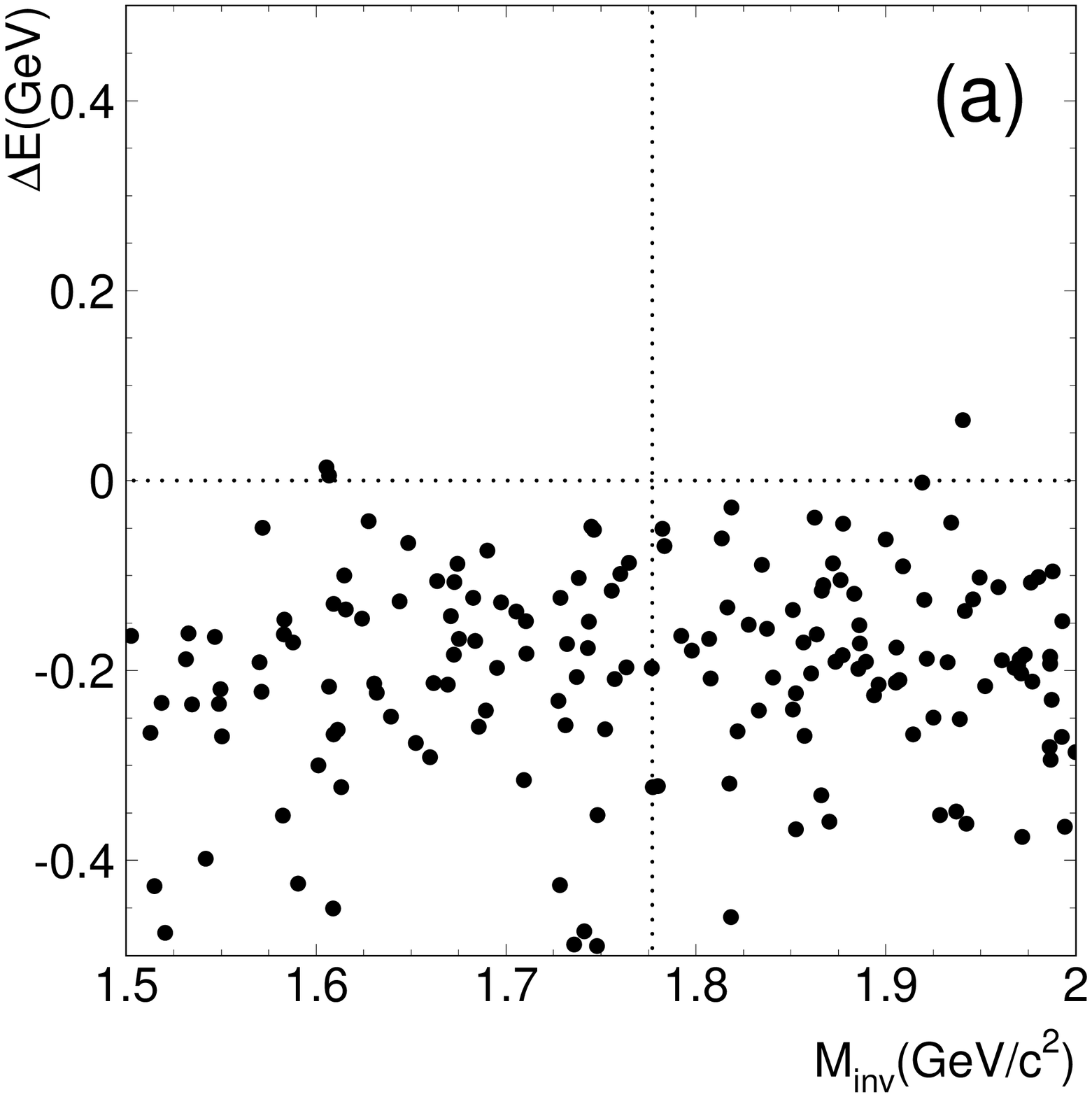}}~~~~~~~~
            \resizebox{5cm}{5cm}{\includegraphics{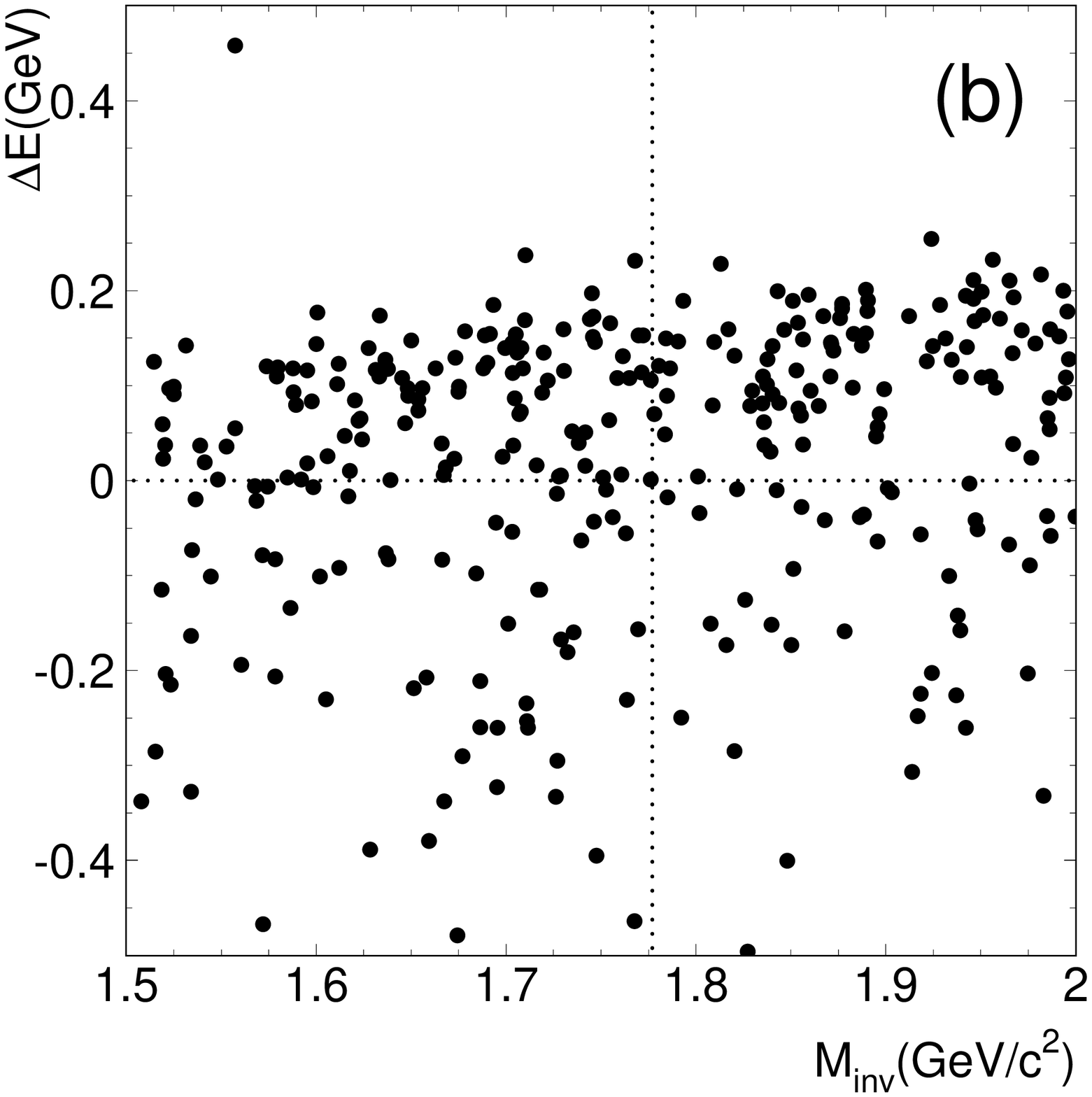}}}
\caption{$M_{\rm inv}$-vs-$\Delta E$ plot of (a) the $\tau$ pair background MC
(140M) and for (b) $N^{\mu\mu}{\rm (data)}$ (32.6 fb$^{-1}$).
The $N^{\mu\mu}{\rm (data)}$ component contributes to the background
through the muon identification inefficiency of about 8\%.}
\label{2d_background}
\end{figure*}

The backgrounds from $\tau$ pairs distribute in the $\Delta E<0$ region,
as shown in Fig.~\ref{2d_background}, and 
could possibly contaminate the signal region.
While a negligibly small portion of $\tau$ pairs is found at $\Delta E>0$,
the data extend to $0<\Delta E< 0.2$.
Our study finds these to be radiative $\mu\mu$ events,
one in which muon is not identified
due to the KLM inefficiency.
Fig.~\ref{2d_background}(b) shows the events that passed
the same selection criteria as the signal,
with the tagged track required to be a muon. We call these ``$\mu\mu$'' events.
The muon identification inefficiency is estimated to be about 8\%.

The background rate is therefore calculated using
\begin{eqnarray}
& & \!\!\!\!\!\!\!\!\!\!\!\!
N_{\rm BG}= 
N^{\mu^{\rm not}\mu}{\rm (MC)}
 \nonumber \\
& & 
+ \{ N^{\mu\mu}{\rm (data)}-N^{\mu\mu}{\rm (MC)} \} \times
\left( \frac{\eta}{1-\eta} \right),
\label{eq1}
\end{eqnarray}
where $N^{\mu^{\rm not}\mu}{\rm (MC)}$ is the number of ``$\mu^{\rm not}\mu$''
events estimated by MC.
The background is mostly $\tau$-pair background and a small portion of the
continuum.
$N^{\mu\mu}{\rm (data)}$ is the number of ``$\mu\mu$'' events 
extracted from data.
The last term is included to avoid double counting. 
The resultant background rates of the individual contributions 
in Eq.(\ref{eq1}) are listed in Table~\ref{summary}.
\begin{table*}[t]
\begin{center}
\caption{Number of the observed events and the expected backgrounds 
for individual components of Eq.(\ref{eq1}) for 
a luminosity of 32.6 $fb^{-1}$.
	}
\label{summary}
\begin{tabular}{|c||c||c|c|c|c||c|}
\hline
 & \multicolumn{1}{c||}{Observed} & 
	\multicolumn{2}{c|}{$N^{\mu^{\rm not}\mu}$} & 
	\multicolumn{2}{c||}{$N^{\mu\mu}$} &
	\multicolumn{1}{c|}{$N_{\rm BG}$} \\
 \cline{3-6}
  	& events & $\tau$ pair & continuum & data 
	& $\tau$ pair &  \\
\hline
Area       & 69 & 34.3 & 5 & 24.4 & 1.2  & 
	62.5$\pm$3.5  \\ 
\hline
\end{tabular}
\end{center}
\end{table*}
Only 5 continuum events (8\% in the Area)
remain in ``$\mu^{\rm not}\mu$'' samples
for 32.6 fb$^{-1}$ luminosity in the Area, but outside of the signal region.
The contribution of the fourth term 
is not significant. It is only 1.9\% in the Area.

\begin{figure}
\centerline{\resizebox{7cm}{4.5cm}{\includegraphics{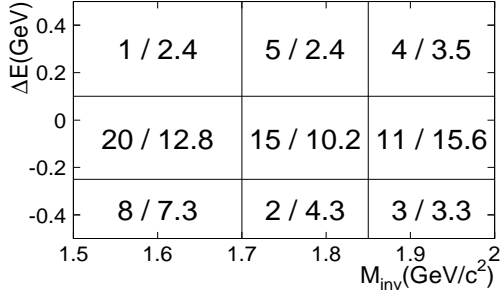}}}
\caption{Number of observed events (left) and 
the expected backgrounds (right) in each region.}
\label{Nbg}
\end{figure}

A comparison between the number of the observed events and 
the expected backgrounds is given in Fig.~\ref{Nbg}. 
The center box is the signal region used in the previous analysis.
We find a good agreement between them.
We observe 69 events with an expected background of $62.5\pm3.5$ events.

\begin{figure*}[t]
\centerline{\resizebox{5cm}{4cm}{\includegraphics{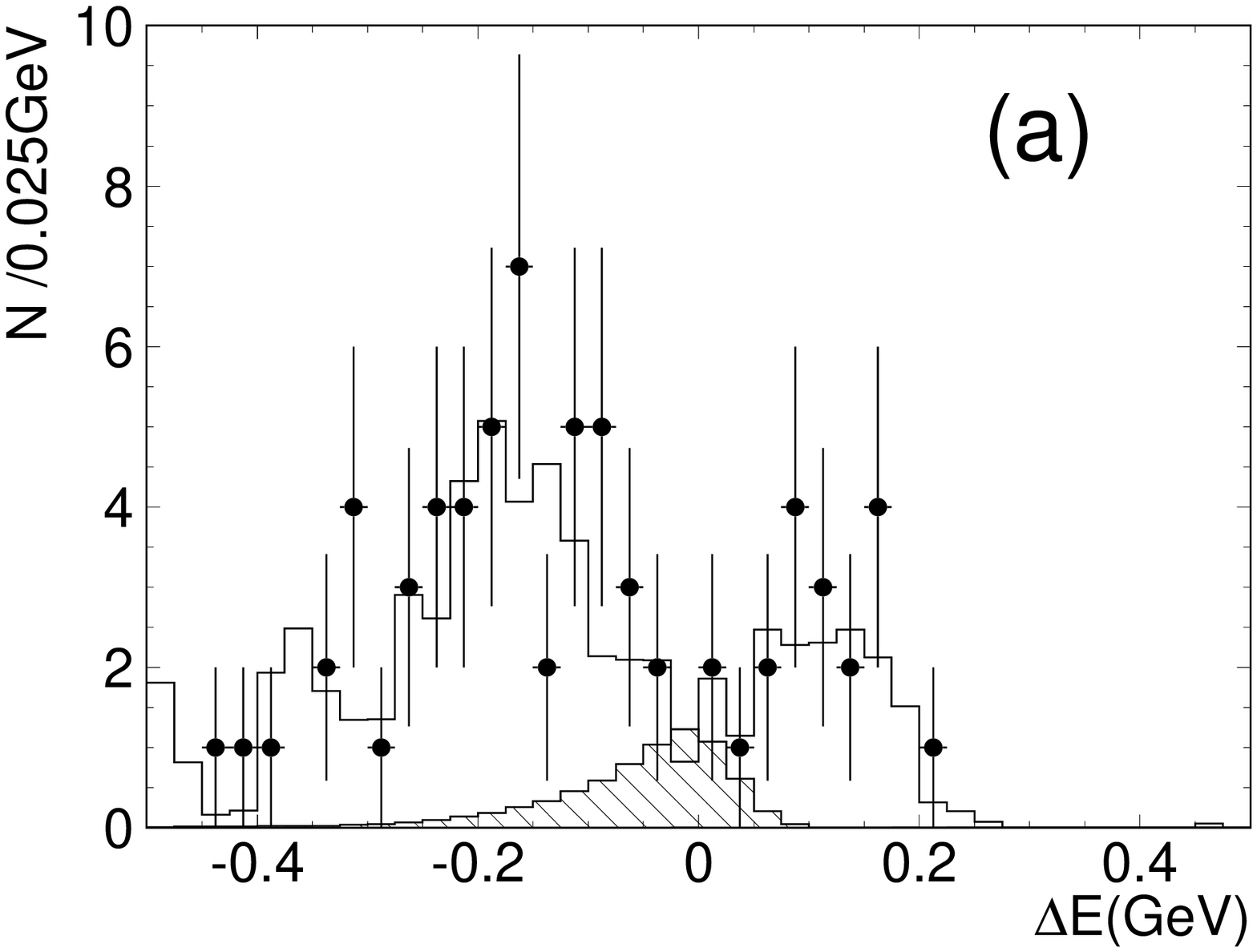}}~~~~~~~~
            \resizebox{5cm}{4cm}{\includegraphics{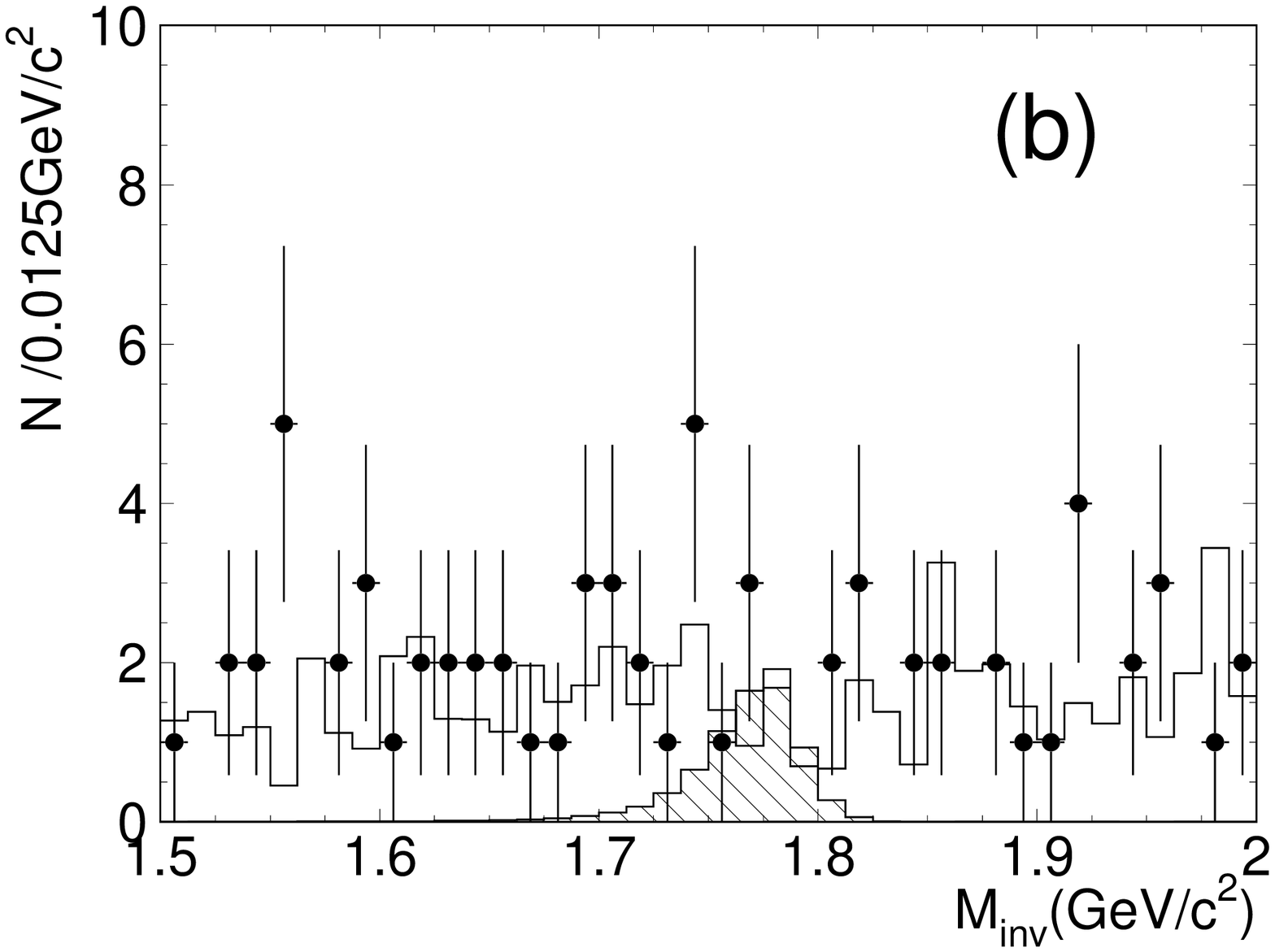}}}
\caption{(a) $\Delta E$ and (b) $M_{\rm inv}$ distributions 
for the surviving events in the Area.
The points with error bars are data,
the open histogram is the expected background evaluated from Eq.(\ref{eq1}),
and the hatched histogram is the signal MC 
(Br($\tau\to\mu\gamma$)=$2\times10^{-6}$).}
\label{de.and.m}
\end{figure*}

\begin{figure*}[t]
\centerline{\resizebox{5cm}{4cm}{\includegraphics{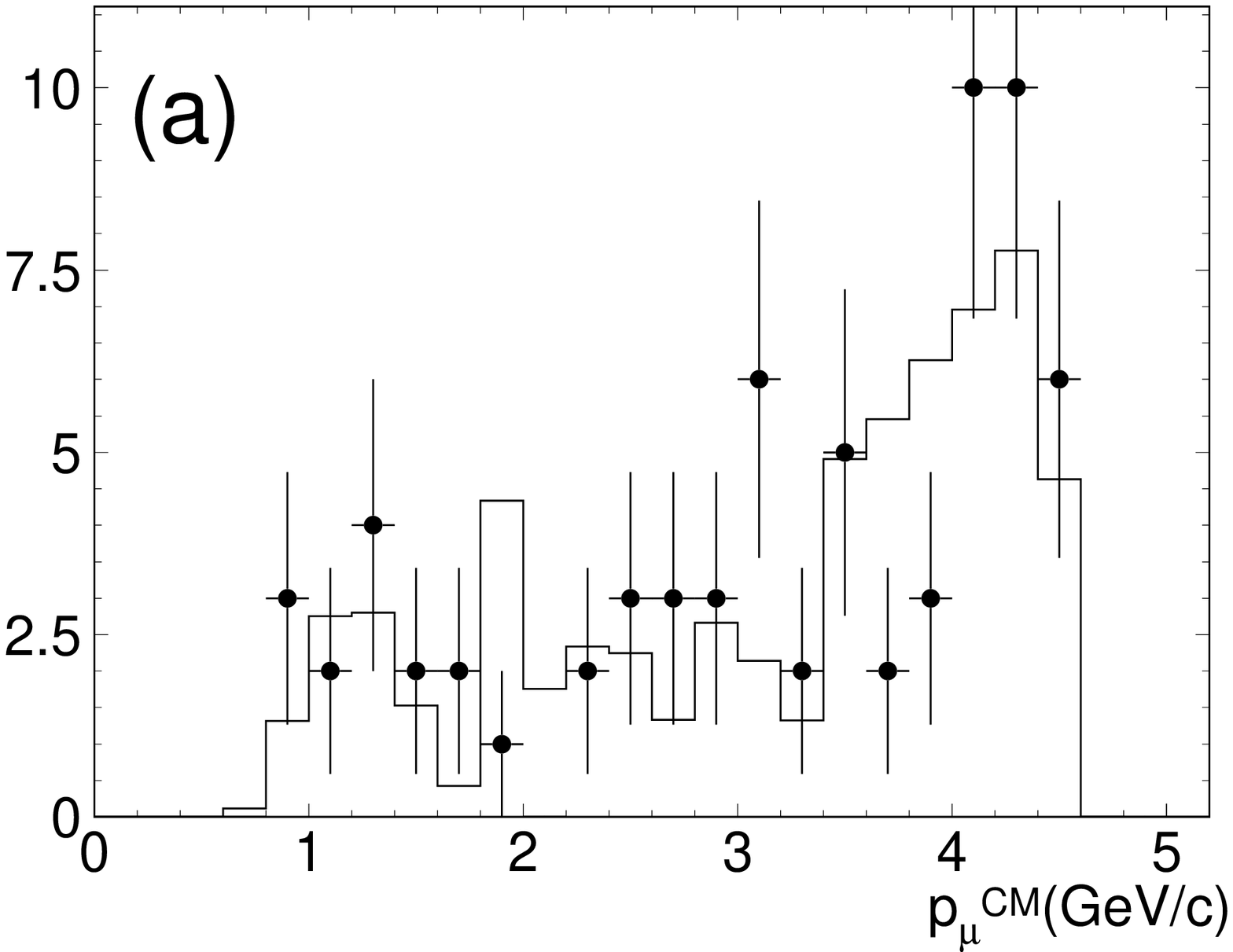}}~~~~~~~~
            \resizebox{5cm}{4cm}{\includegraphics{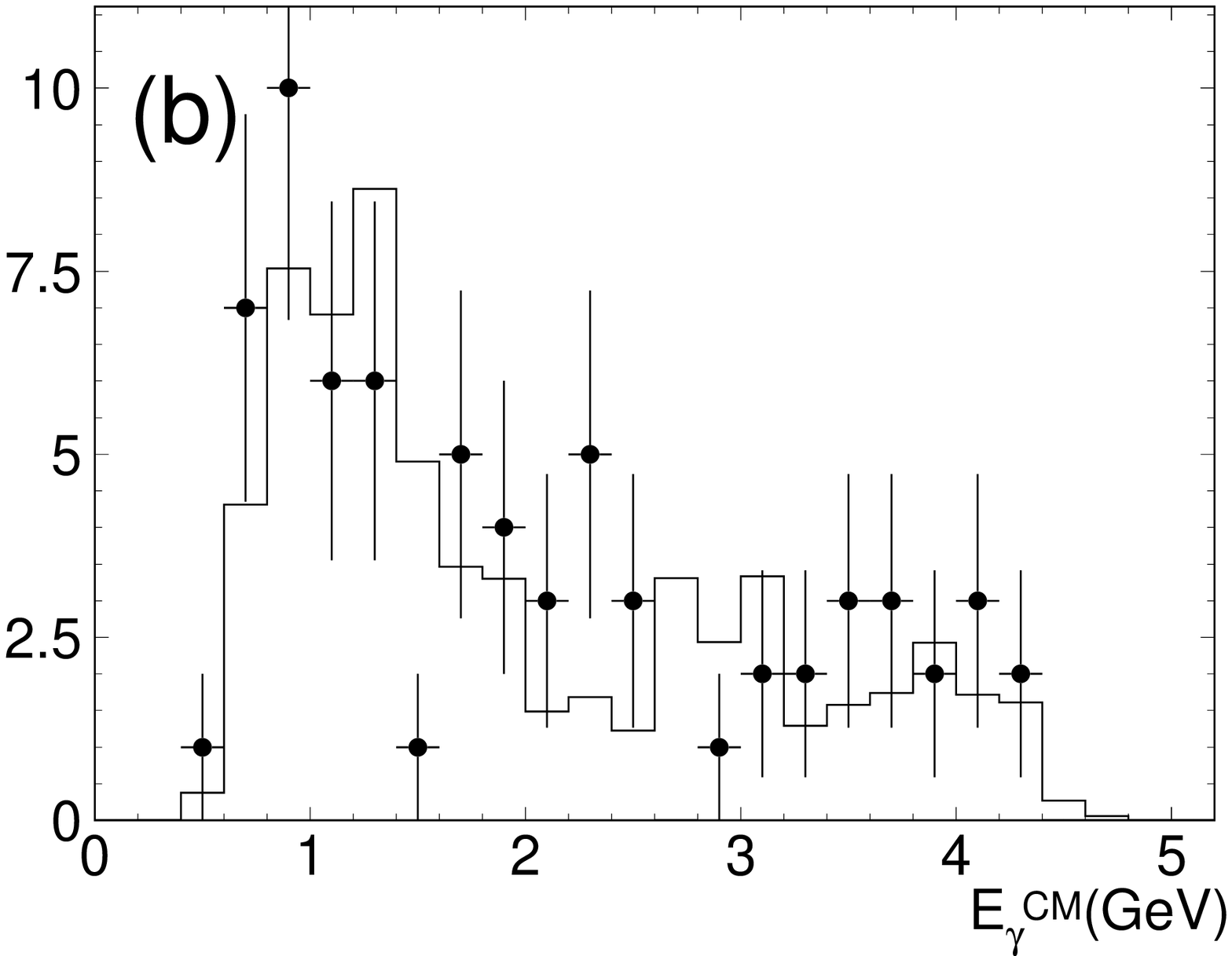}}}
\caption{(a) $p_\mu^{\rm CM}$ and (b) $E_\gamma^{\rm CM}$ distributions
of the events that survived all selections.
The points with error bars are data and
the open histogram is the expected background evaluated from Eq.(\ref{eq1}).}
\label{pm.and.eg}
\end{figure*}

Comparisons of the $\Delta E$ and $M_{\rm inv}$ distributions
between data (dots) and the expected background (open histogram)
are also shown in Fig.~\ref{de.and.m}.
While a uniform distribution is seen for $M_{\rm inv}$,
a two-bump structure is found in the $\Delta E$ distribution.
The two bumps corresponds to $\tau$ pair backgrounds and radiative $\mu\mu$
processes.
Fig.~\ref{pm.and.eg} shows the $p_\mu$ and $E_\gamma$
distributions of the signal candidate 
for the same samples of Fig.~\ref{de.and.m}.
Both bumps in the high-$p_\mu$ and the low-$E_\gamma$ regions
are due to the remaining radiative $\mu\mu$ samples.
As can be seen in Figs.~\ref{de.and.m} and \ref{pm.and.eg},
the distributions of the expected backgrounds are 
very consistent with the data.

\section{Result}

In the signal region, we find one event, as shown in Fig.~\ref{2d_data}.
The number of background events is evaluated to be $2.5 \pm 0.6$.
The upper limit is obtained using the Bayesian approach 
(following refs \cite{CLEO}, \cite{Helene} and \cite{PDG})
with the equation
\begin{equation}
\frac{
e^{-(s_0+b_0)} \sum_{n=0}^{n_0}(s_0+b_0)^n/n! }
{e^{-b_0} \sum_{n=0}^{n_0}b_0^n/n!} = 0.1,
\end{equation}
where $s_0$ is the upper limit on the signal at 90\% 
confidence, $b_0$ is the number of expected background events
and $n_0$ is the number of observed events.
We obtain $s_0 = 4.1$.

The systematic uncertainties of the detection efficiency are listed 
in Table~\ref{system}.
The conservative and preliminary evaluation gives a 6.7\% error in total.
\begin{table}
\caption{Systematic uncertainties in the detection efficiency.
The total uncertainty is evaluated by adding them in quadrature. } 
\label{system}
\begin{center}
\begin{tabular}{|l|c|}
\hline
 &uncertainty (\%) \\
\hline
track rec. eff. & 2.0\\
photon rec. eff. & 5.0\\
cut selection & 2.2 \\
luminosity & 1.4 \\
muon identification & 2.3  \\ 
MC statistics & 0.8 \\
trigger efficiency & 1.6 \\
\hline
total & 6.7 \\
\hline
\end{tabular}
\end{center}
\end{table}

Finally,
we obtain a preliminary result for the upper limit of Br($\tau \to \mu \gamma$)
using 29.7 million $\tau$ pairs
with $\epsilon=$9.0\% detection efficiency, as
\[
Br(\tau \to \mu \gamma)< 6 \times 10^{-7}. 
\]


\begin{thebibliography}{99}
\bibitem{SUSY}{R. Kitano and Y. Okada, Phys. Rev. {\bf D63} 
	(2001) 113003; S.F. King, Phys. Rev. {\bf D60} (1999) 035003; 
	J. Hisano and D. Nomura, Phys. Rev. {\bf D59} (1999) 116005; 
	K.S. Babu, B. Dutta, and R.N. Mohapatra, Phys. Lett. 
	{\bf B458} (1999) 93; 
	J. Hisano, T. Moroi, K. Tobe, M. Yamaguchi, and T. Yanagida, 
	Phys. Lett. {\bf B357} (1997) 579; 
	S. Gentile and M. Pohl, Phys. Report 274 (1996) 287. }
\bibitem{CLEO}{K.W. Edwards et al. (CLEO Collaboration), 
	Phys. Rev. {\bf D55}, (1997) 3919; 
	S. Ahmed et al., Phys. Rev. {\bf D61} (2000) 071101.} 
\bibitem{EPS01}{\em {Search for $\tau \to \mu \gamma$ decay at Belle}}, 
	Proc. of Int. Europhy. Conf. on HEP, July 12-18, 2001, 
	Budapest, Hungary.
\bibitem{ICHEP02}{\em {Search for $\tau \to \mu \gamma$ at BaBar}}, 
	Proc. of XXXI Int. Conf. on HEP, July 24-31, 2002, 
	Amsterdam, the Netherlands.
\bibitem{Belle} A.~Abashian et al. (Belle Collaboration),
        Nucl. Instr. and Meth. A {\bf 479} (2002) 117.
\bibitem{KEKB} E.~Kikutani ed., KEK Preprint 2001-157 (2001), 
         to appear in Nucl. Instr. and Meth. A.
\bibitem{KORALB}{J. Jadach and Z. Was, Comp. Phys. Commun. 
	{\bf 85} (1995) 453.} 
\bibitem{QQ} 'QQ - The CLEO Event Generator' (http:// \\
	www.lns.cornell.edu/public/CLEO/soft/QQ).
\bibitem{GEANT}{R. Brun et al., GEANT 3.21, 
	CERN Report No. DD/EE/84-1 (1987). }
\bibitem{Helene}{O. Helene, Nucl. Instr. and Meth. {\bf 212} (1983) 319.}
\bibitem{PDG}{Particle Data Group, R.M. Barnett et al., 
	Phys. Rev. {\bf D54}, 1 (1996), p. 159.}
\end{thebibliography}
\end{document}